\documentclass[letterpaper, 10 pt, conference]{ieeeconf}  % Comment this line out if you need a4paper

\IEEEoverridecommandlockouts                              % This command is only needed if 
\usepackage[pdftex]{graphicx}
\graphicspath{{fig/}}
\usepackage{amsmath}
\usepackage{amssymb}
\usepackage{amsfonts}
\usepackage{gensymb}
\usepackage{mathtools}
\usepackage{multirow}
\usepackage{comment}
\usepackage{color}
\usepackage{array}
\usepackage{url}
\usepackage{hyperref}
\hypersetup{colorlinks=true, linkcolor=black, filecolor=black,   urlcolor=black,citecolor=black, anchorcolor=cyan}
\usepackage{cite}
\usepackage[strings]{underscore}
\let\labelindent\relax
\usepackage{enumitem}
\usepackage{soul}
\usepackage{subcaption}
\usepackage{algorithm2e}
\SetKwComment{Comment}{/* }{ */}
\RestyleAlgo{ruled}

\renewcommand{\theenumi}{\roman{enumi}}

\newcommand{\It}{\ensuremath{I_{3\times 3}}}
\newcommand{\zerot}{\ensuremath{0_{3\times 3}}}
\newcommand{\R}{\mathbb{R}}

\newcommand{\bk}{\color{black}}

\title{\LARGE \bf Cooperative Bearing-Only Target Pursuit via Multiagent  \\ Reinforcement Learning: Design and Experiment}

\author{
	Jianan~Li,
	Zhikun~Wang,
	Susheng~Ding,
	Shiliang~Guo,
	Shiyu~Zhao
	\thanks{* To appear in the 2025 IEEE/RSJ International Conference on Intelligent Robots and Systems (IROS 2025).}
	\thanks{This research work was supported by the STI 2030-Major Projects (Grant No. 2022ZD0208800) and National Natural Science Foundation of China (Grant No. 62473320).}
	\thanks{J. Li is with the Shanghai AI Laboratory, 200030, Shanghai, China and WINDY Lab in the Department of Artificial Intelligence at Westlake University, 310024, Hangzhou, China (E-mail: lijianan@westlake.edu.cn).}
	\thanks{Z. Wang*, S. Ding, S. Guo and S. Zhao are with WINDY Lab in the Department of Artificial Intelligence at Westlake University, 310024, Hangzhou, China (E-mail: \{wangzhikun, dingsusheng, guoshiliang, zhaoshiyu\}@westlake.edu.cn). *Corresponding author.} 
}

\begin{document}

\maketitle
\thispagestyle{empty}
\pagestyle{empty}

%%%%%%%%%%%%%%%%%%%%%%%%%%%%%%%%%%%%%%%%%%%%%%%%%%%%%%%%%%%%%%%%%%%%%%%%%%%%%%%%
\begin{abstract}
	This paper addresses the multi-robot pursuit problem for an unknown target, encompassing both target state estimation and pursuit control. 
	First, in state estimation, we focus on using only bearing information, as it is readily available from vision sensors and effective for small, distant targets. 
	Challenges such as instability due to the nonlinearity of bearing measurements and singularities in the two-angle representation are addressed through a proposed uniform bearing-only information filter. This filter integrates multiple 3D bearing measurements, provides a concise formulation, and enhances stability and resilience to target loss caused by limited field of view (FoV).
	Second, in target pursuit control within complex environments, where challenges such as heterogeneity and limited FoV arise, conventional methods like differential games or Voronoi partitioning often prove inadequate. To address these limitations, we propose a novel multiagent reinforcement learning (MARL) framework, enabling multiple heterogeneous vehicles to search, localize, and follow a target while effectively handling those challenges.
	Third, to bridge the sim-to-real gap, we propose two key techniques: incorporating adjustable low-level control gains in training to replicate the dynamics of real-world autonomous ground vehicles (AGVs), and proposing spectral-normalized RL algorithms to enhance policy smoothness and robustness.
	Finally, we demonstrate the successful zero-shot transfer of the MARL controllers to AGVs, validating the effectiveness and practical feasibility of our approach. 
	The accompanying video is available at https://youtu.be/HO7FJyZiJ3E.
\end{abstract}

\section{Introduction}

\begin{figure}[t]
	\centering
	\includegraphics[width=1\linewidth]{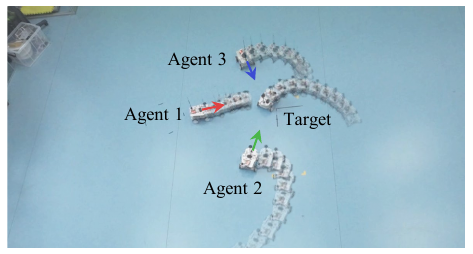}
	\caption{Trajectories of three pursuer vehicles and the target in one of our experiments. Arrows represent the AGVs' headings, aligned with the center lines of their FoV.}
	\label{fig:demo}
\end{figure}

The pursuit-evasion game has been a subject of significant research focus over several decades. It serves as a fundamental yet versatile mathematical framework with diverse practical applications, including autonomous guidance \cite{faruqi2017differential}, criminal pursuit \cite{diestel2024graph}, aerospace \cite{pontani2009numerical} and robotics \cite{bajcsy2024learning}. 
In a multiagent pursuit-evasion game, an extension of the classic one-on-one pursuit-evasion scenario, multiple pursuers coordinate to capture or intercept evaders within a given environment (see Fig.~\ref{fig:demo}).
Understanding and analyzing multiagent pursuit-evasion games provide valuable insights into cooperative strategic decision-making \cite{li2023predator}.
However, solving multiagent pursuit-evasion games is a complex challenge due to unknown target states, dynamic interactions and multiple constraints among agents, resulting in a high-dimensional, non-linear, and often non-convex problem.

The \textit{first step} to realize unknown target pursuit is to estimate target state. Vision-based solutions serve as a prominent method for target perception. However, the readily available information from visual sensors is target bearing, and depth can not be recovered directly when the target size is unknown \cite{tekin2018real,vrba2020marker, li2021image}. 
Although SLAM-based methods offer a potential solution, they rely on feature points extraction or 3D bounding boxes for state estimation \cite{yang2019cubeslam} \cite{qiu2019tracking}. These approaches become ineffective when the target is distant and small, as insufficient stable features can be extracted, and 3D detection is not feasible \cite{ning2024bearing}.
These limitations restrict us to addressing the bearing-only target state estimation problem \cite{li2022three,ning2024bearing}. 
A notable challenge in bearing-only state estimation lies in the inherent non-linearity between the measurement and target state, which leads to biased or even divergent estimation outcomes for conventional methods like the extended Kalman filter. Although the utilization of modified polar or spherical coordinates\cite{stallard1991angle, van2021state}, and the pseudo-linear Kalman filter (PLKF) \cite{yang2023novel, aidala1982biased, he2018three} alleviates the divergence problem, most of them  are analyzed in 2D scenarios. When applied to 3D cases, the two-angle representation including azimuth and elevation introduces singularities at an elevation of 90$\degree$.

The \textit{second step} is target pursuit control.
While researchers have made significant progress in understanding and addressing specific instances of pursuit-evasion games, much of the current research focuses on relatively idealized scenarios. 
For instance, the widely used differential game framework typically assumes a point-mass model with constant velocities and the number of agents is limited \cite{isaacs1999differential, faruqi2017differential}. As the number of agents increases, solving the minimax optimization problem using Hamilton-Jacobi-Issacs partial differential equations becomes intractable \cite{zhou2016cooperative}. Similarly, the Voronoi partition methods further assume the agent positions and boundary information are fully known \cite{huang2011guaranteed,zhou2016cooperative}.

Recent research has embraced MARL as a promising approach. This is primarily attributed to its model-free nature enabling handling of complex environments and adaptability \cite{zhao2024reinforcement}.
However, the majority of existing literature primarily focuses on enhancing the learning efficiency of MARL algorithms \cite{gupta2017cooperative, lowe2017multi}, rather than the specific pursuit-evasion games where practical challenges remain unaddressed. For example, the target position estimate and the FoV constraint are usually neglected. 
Furthermore, other research focuses on various aspects of multiagent systems, such as tackling the permutation invariance problem \cite{huttenrauch2019deep} or managing the communication load \cite{wang2020cooperative}. 
It is worth noting that a significant portion of these studies relies solely on simulation\cite{zhou2021multi,qu2023pursuit}, lacking experimental validation.

In this work, we aim to provide a comprehensive solution for an unknown target pursuit by a group of heterogeneous robots using only bearing information.
The contribution and novelty of this paper are summarized as follows.

\begin{enumerate}[wide, nolistsep, label={\arabic*)}]
	\item 
	We propose a novel cooperative uniform bearing-only pseudo-linear information filter. It integrates multiple bearing measurements in 3D, removes the traditional singularity problem, and offers a concise yet effective formulation.
	\item  
	We propose a novel MARL framework for designing a multi-robot pursuit controller that enables heterogeneous vehicles to search, localize, and follow a target in complex environments while addressing constraints such as kinematics, FoV, range, and observability, which is challenging to resolve concurrently using conventional methods.
	\item 
	To bridge the sim-to-real gap, we propose two techniques: introducing adjustable low-level control gains during training to mirror real-world AGV control responses, and spectral-normalized actor-critic RL algorithms to improve the smoothness and robustness of learned policies. Extensive experiment results validate their effectiveness.
\end{enumerate}

\section{Problem Statement and Overview} \label{sec:problemStatement}

Consider a target moving with unknown position and velocity in space. Multiple heterogeneous pursuer vehicles, capable of omnidirectional or unicycle movement, use onboard vision to detect the target and hence obtain their relative bearings. 
The primary objective of the pursuer vehicles is to search for, localize, and then follow the target from any initial distant location and heading angles. 
Simultaneously, the pursuers must satisfy motion, FoV, observability, collision avoidance, and range maintenance constraints.

\begin{figure}[t]
	\centering
	\includegraphics[width=0.995\linewidth]{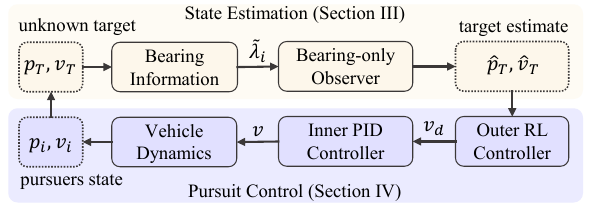}
	\caption{Architecture of the proposed approach.}
	\label{fig:architecture}
\end{figure}

In order to achieve this goal, we design a compound system (see Fig.~\ref{fig:architecture}). 
Let $p_T,v_T\in\R^3$ be the position and velocity of the target, and $p_i,v_i\in\R^3, i=\{1,2,...,n\}$, the position and velocity of the $i$th pursuer vehicle, all expressed in an inertial coordinate frame. 
The outputs of multiple vision systems are noise-corrupted bearing measurements $\tilde{\lambda}_i$. 
The first step is to design a bearing-only observer to estimate the target position and velocity (see Section \ref{sec:estimation}). 
Building upon the target estimate, we proceed to design a pursuit controller based on MARL (see Section \ref{sec:rlFramework}).

\section{Target State Estimation} \label{sec:estimation}

Effective target pursuit requires first estimating the target state. In this section, we extend the work in \cite{li2022three}, and introduce a novel 3D bearing-only cooperative pseudo-linear information filter (PLIF) to estimate the target position and velocity from multiple noisy bearing measurements.

\subsection{State Transition Equation} 
Denote the target state as $x_T=[p_T^T,v_T^T]^T \in \mathbb{R}^6$, and $i$th pursuer state as $x_{Pi}=[p_i^T,v_i^T]^T \in \mathbb{R}^6$. Note $x_T$ is to be estimated and $x_{Pi}$ is known. The states propagation are described by discrete double-integrator dynamics:
\begin{subequations}
	\begin{align}
		x_{T,k} &= Ax_{T,k-1} + Bq_k \label{eq:dotxT} \\
		x_{Pi,k} &= Ax_{Pi,k-1} + Bu_{i,k} \label{eq:dotxM}
	\end{align}
\end{subequations}
where
\begin{equation*}
	A = \begin{bmatrix}
		I_{3\times 3} & \Delta t\It \\
		\zerot & \It
	\end{bmatrix},
	B = \begin{bmatrix}
		\frac{1}{2}\Delta t^2 \It \\
		\Delta t \It
	\end{bmatrix},
\end{equation*}
and $\Delta t$ is the sampling time step.
Here, $u_{i,k}\in \mathbb{R}^3$ is the control input of the $i$th pursuer, $q_k$ is a process noise drawn from a zero-mean \bk normal distribution, that is,
$q_k \sim \mathcal{N}(0,Q_k)$ where we assume the acceleration of the target is unknown.

\subsection{Measurement Equation}
Instead of by using azimuth and elevation angles, we use a vector to represent the measured bearing information:
\begin{equation}
	\tilde{\lambda}_i = \lambda_i + \nu_i =  \frac{p_T - p_i}{||p_T - p_i||} + \nu_i  \label{eq:lambda0}
\end{equation}
where $\lambda_i$ is the true bearing vector for $i$th pursuer, $\nu_i \in \mathbb{R}^3$ is the measurement noise assumed to be normally distributed: $\nu_i \sim \mathcal{N}(0, \Sigma), \Sigma = \sigma^2 \It$.
By adopting the vector representation, we can eliminate the need for complex trigonometric functions and mitigate the risk of encountering singularities, thus ensuring the robustness of the filter.

Inspecting  \eqref{eq:lambda0}, we notice that $\tilde{\lambda}_i$ is a nonlinear function of $p_T$. To transform the nonlinear measurement transition matrix into a linear form, we introduce an orthogonal projection operator inspired by our previous work \cite{zhao2019bearing}. For any nonzero vector $g\in \R^3$, the orthogonal project operator is defined as
\begin{equation}
	P_g = I_{3\times3} - \frac{g}{\|g\|}\frac{g^T}{\|g\|} \label{eq:proj}.
\end{equation}
To interpret $P_g$, consider a vector $z \in \mathbb{R}^3$. The orthogonal projection of $z$ onto the plane perpendicular to $g$ is $P_g z$.

Multiplying $r_iP_{\tilde{\lambda}_i}$ on both sides of \eqref{eq:lambda0} and rearrange gives
\begin{equation} \label{eq:meaEqSimple}
	P_{\tilde{\lambda}_i}p_i = P_{\tilde{\lambda}_i} p_T + r_iP_{\tilde{\lambda}_i}\nu_i
\end{equation}
where $r_i=||p_T-p_i||$ and the fact $P_{\tilde{\lambda}}\tilde{\lambda} = 0$ is used.
Thus, the nonlinear measurement function becomes a pseudo-linear one.
Different to the work in \cite{li2022three}, the state vector in Eq.~\eqref{eq:meaEqSimple} becomes the target position $p_T$ in the inertia coordinate rather than the relative position $p_T - p_i$. The rationale behind this stems from the necessity to assimilate multiple measurements.
Augmenting the state vector to include both position and velocity, the measurement equation is finally obtained as
\begin{equation} \label{eq:meaEq}
	\begin{bmatrix}
		P_{\tilde{\lambda}_i} &  0_{3\times 3}
	\end{bmatrix} x_{Pi}
	= \begin{bmatrix}
		P_{\tilde{\lambda}_i} &  0_{3\times 3}
	\end{bmatrix} x_T + r_iP_{\tilde{\lambda}_i}\nu_i.
\end{equation}

\subsection{3D Bearing-Only Cooperative PLIF}
In the case of multiple measurements, the measurement matrix in Eq.~\eqref{eq:meaEq} needs to be augmented. When the number of measurements is large, it is computationally expensive to inverse the large matrix in the computation of Kalman gain. To address this issue, we propose to transform the Kalman form into an information form. 
This approach also ensures that the estimation process remains robust when the number of measurements fluctuates or even drops to zero due to limited FoV. Our experiment confirms its flexibility and robustness (see Fig.~\ref{fig:experiment results}ef).

First, we replace the covariance matrix by its inverse, denoted as $Y_k$ at time step $k$. The state vector is then substituted by the information vector $y_{T,k}=Y_k x_{T,k}$. 
For simplicity, we will omit the subscript $T$ in the following expressions, thus, $y_{k}=Y_k x_{k}$. 
Employing transformations using the Sherman-Morrison-Woodbury identity for the inverse of a matrix, the information form can be derived. The steps of the proposed 3D bearing-only observer under the framework of an information filter are summarized below. The prediction step is \bk
\begin{align*}
	M_{k-1} &= (A^{-1})^T Y_{k-1} A^{-1} \\
	Y_k^- &= (I_{6\times 6} + M_{k-1}BQ_kB^T)^{-1} M_{k-1} \\
	\hat y_k^- &= (I_{6\times 6} + M_{k-1}BQ_kB^T)^{-1} (A^T)^{-1} \hat y_{k-1}
\end{align*}
where $	\hat y_k^-$ and $Y_k^-$ are the \textit{prior} information state estimate and information matrix. 
The correction step \bk is
\begin{align*}
	\hat y_k &= \hat y_k^- + \sum_{i=1}^{n} H_{i,k}^T (V_{i,k}\Sigma_i V_{i,k}^T)^{\dagger} H_{i,k}x_{Pi} \\
	Y_k &= Y_k^- + \sum_{i=1}^{n} H_{i,k}^T (V_{i,k}\Sigma_i V_{i,k}^T)^{\dagger}H_{i,k}
\end{align*}
where $H_{i,k} = \begin{bmatrix}
	P_{\tilde{\lambda}_{i,k}} &  0_{3\times 3}
\end{bmatrix}$, 
$V_{i,k}=\hat{r}_{i,k}^-P_{\tilde{\lambda}_{i,k}}$;
$\hat{r}_{i,k}^-$ is simply the norm of the first three components of $\hat{x}_{i,k}^-$; 
$\hat{y}_k$ and $Y_k$ are the \bk \textit{posterior} information state estimate and information matrix.
The reason for using pseudo-inverse $\dagger$ is that the matrix $(V_{i,k}\Sigma_i V_{i,k}^T)$ is rank deficient since $P_{\tilde{\lambda}_{i,k}}$ in $H_{i,k}$ is rank deficient.
This is a common practice when the inverse does not exist \cite{anderson2012optimal, yoshikawa1972discrete,kulikov2018moore}.
Following the naming convention in \cite{lingren1978position, li2022three}, The proposed cooperative uniform pseudo-linear information filter is coined as \textit{u-PLIF}.

\section{Target Pursuit Control} \label{sec:rlFramework}

After estimating the target state, the next step is to achieve target pursuit control. In this section, we propose a MARL framework for developing cooperative pursuit controllers.

\begin{figure}[t]
	\centering
	\includegraphics[width=0.99\linewidth]{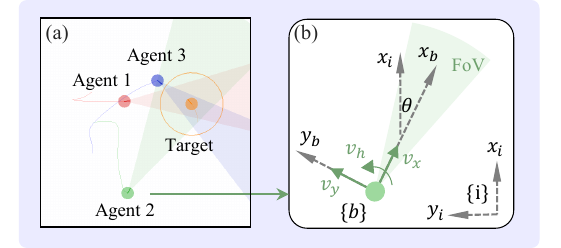}
	\caption{(a) Training environment. (b) Agent actions and FoV.}
	\label{fig:action}
\end{figure}
\subsection{Environment, Agent Dynamics and Action Space} \label{subsec:agent dynamics}
We first developed a physics-based simulation environment for MARL, where pursuer and target vehicles interact within a two-dimensional continuous space \cite{li2023predator}, as shown in Fig.~\ref{fig:action}a. Pursuer agents and their FoV are in red, green and blue, respectively. The target is depicted in orange with a circle representing the control range.
Additional details of an agent are illustrated in Fig.~\ref{fig:action}b.
A body-frame $\{b\}$ is rigidly attached on the agent with its x-axis $x_b$ aligned with the forward direction of the vehicle.
A simulated forward-looking camera with limited FoV is assumed to be installed on the agent aligned with the x-axis of the body frame.
An agent is subject to three actions:  the angular velocity of the heading $v_h \in \mathbb{R}$, and the linear velocity commands $v_x, v_y \in \mathbb{R}$ in the x-, y-axis of $\{b\}$. To summarize, the actions of the agent are $[v_h,v_x, v_y]^T \in \mathbb{R}^3$.
% $[\omega_d,v_{dx}, v_{dy}]

The dynamics of an agent $i$ are modeled as second-order integrator dynamics:
\begin{subequations} \label{eq:dynamics}
	\begin{align}
		\dot p &= v                     \label{eq:dynamics:p}\\
		\dot v &= a + (f_a + f_b) / m   \label{eq:dynamics:v}   \\
		\dot R & = R[\omega]_\times     \label{eq:dynamics:R}    \\
		\dot \omega &= \alpha           \label{eq:dynamics:w}
	\end{align}
\end{subequations}
where Eq.~\eqref{eq:dynamics:p} and \eqref{eq:dynamics:v} describe the translational dynamics, $p \in \mathbb{R}^2$ is the position, $v\in \mathbb{R}^2$ is the velocity, $a \in \mathbb{R}^2$ is the linear acceleration control input, $f_a \in \mathbb{R}^2$ is the elastic force between contact agents, $f_b \in \mathbb{R}^2$ is the elastic force between agents and boundaries. Elastic forces reflect physical collision dynamics following Hooke's law, and sum up when an agent contacts multiple other agents or boundaries as $f_a=\sum_j f_{a,j}$ and $f_b=\sum_j f_{b,j}$, and $m \in \mathbb{R}^+$ is the mass.
Equations \eqref{eq:dynamics:R} and \eqref{eq:dynamics:w} describe the rotational dynamics, where $R \in \mathbb{R}^{2\times 2}$ is the rotation matrix from the vehicle's body frame $\{b\}$ to the fixed inertial frame $\{i\}$, $\omega \in \mathbb{R}$ is the angular speed. Specifically, 
\begin{equation*}
	R = \begin{bmatrix}
		\cos\theta & -\sin \theta \\
		\sin\theta & \cos \theta \\
	\end{bmatrix}, \quad
	[\omega]_\times = \begin{bmatrix}
		0 & -\omega \\
		\omega & 0 
	\end{bmatrix},
\end{equation*}
where $\theta$ is the heading angle, and $\alpha\in \mathbb{R}$ is angular acceleration control input.

\subsection{Observation Space and Reward Design}

The observation vector for a pursuer agent is composed of three components: ego-state, allies' state, and target state, providing a comprehensive environmental understanding for decision-making.
The ego-state includes the agent's position, velocity, and heading. Allies' states, shared via communication, include relative position, velocity, heading, and a binary flag indicating target detection. The target state is available only if detected within the pursuer's limited FoV; otherwise, a zero mask is applied.

Aligning with the primary goal and constraints described in Section \ref{sec:problemStatement}, the reward function is $r=\sum_{j=1}^{5} r_j$ including:
$r_1=0.2$, if the agent executes counterclockwise rotation for target searching;
$r_2 = 1$, if $\lambda^Th$ exceeds a threshold for FoV constraint;
$r_3 = 1$, if the range between the target and the closest pursuer falls below a predetermined threshold;
$r_4 = \det(\mathcal{I})$, where $\mathcal{I} = \sum_{i=1}^n (I_{2\times 2}-\lambda \lambda^T)$ for observability enhancement\cite{li2022three};
$r_5 = -10$, if collision happens.

\subsection{Sim-to-Real} \label{sec:spectralNormalization}

Control policies learned in a simulated environment are often challenging to apply directly to real robotic systems due to the simulation-to-reality gap, which arises from differences in dynamics, noise, and unforeseen variables. To mitigate this, we introduce two techniques in our experiment.

\begin{figure*}[t]
	\centering
	\includegraphics[width=1\linewidth]{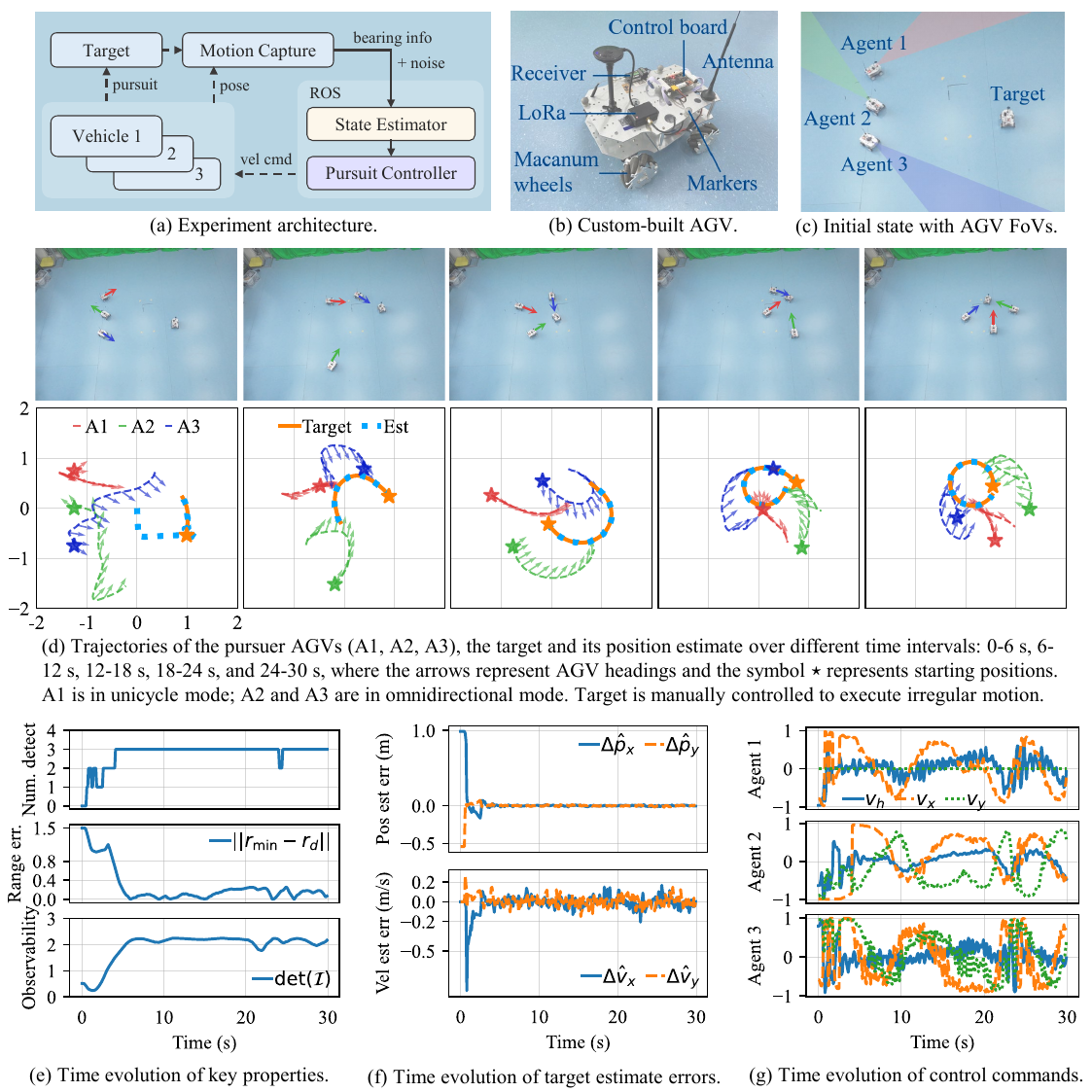}
	\caption{Experiment Result (See the accompanying video at https://youtu.be/HO7FJyZiJ3E.  More original experimental footage is available at https://youtu.be/70ead3rQcok.)}
	\label{fig:experiment results}
\end{figure*}

The \textit{first technique} involves incorporating adjustable low-level control gains during the training process to more accurately replicate the real-world control dynamics of AGVs. The vehicles used in our experiment are Macanum wheeled vehicles which are directly controlled via velocity commands by adjusting the rotary wheel speed (see Fig.~\ref{fig:experiment results}b). In real-world AGVs, there is an inherent acceleration phase required to reach the desired velocity. To simulate this transition behavior in the training environment, we model it as a first-order lag, which can be viewed as a low-level control mechanism. Consequently, the control inputs $a$ and $\alpha$ in Eq.~\eqref{eq:dynamics} are further designed as follows: 
\begin{subequations} \label{eq:acalphac}
	\begin{align}
		a &= k_v(Rv_d - v) \label{eq:acalphac:a}\\
		\alpha &= k_\omega (\omega_d - \omega)
	\end{align}
\end{subequations}
where $v_d =[v_x, v_y]^T$ and $\omega_d=v_h$ are the desired linear and angular velocities, respectively, determined by policy neural networks; $k_v, k_\omega \in\mathbb{R}^+$ are control gains, determined by real robot dynamics. 
We tuned $k_v$ and $k_\omega$ through step response experiments to align the velocity response in training with the real-world behavior of AGVs. Without this adjustment, RL-controlled AGVs tend to exhibit oscillatory behavior when tracking control commands. While not necessarily required for traditional controllers like PID, it is crucial for RL systems due to their increased sensitivity and reliance on simulation accuracy.

It is remarked that by setting the lateral velocity $v_y$ to zero, the vehicle transitions into the unicycle mode. This enables the system to be considered heterogeneous, accommodating both omnidirectional and unicycle modes. Furthermore, the rotation matrix $R$ preceding $v_d$ in Eq.~\eqref{eq:acalphac:a} is essential for converting the control command from the body frame $\{b\}$ to inertial frame $\{i\}$.

The \textit{second technique} involves using spectral-normalized actor networks in RL. During training, uncorrelated random noise is injected into the action space to enhance exploration. However, this results in oscillatory and jerky control signals, causing the vehicle to struggle with tracking rapidly changing commands \cite{shen2020deep,ibarz2021train}. Real-world noise further exacerbates these jerky motions, leading to excessive power consumption and increased wear on the robotic system \cite{mysore2021regularizing}. The primary cause of this issue is the lack of Lipschitz continuity in the policy network \cite{shi2021neural}.

A function $f: \mathbb{R}^m \rightarrow \mathbb{R}^n$ is said to be \textit{Lipschitz continuous} if there exists a constant $L$ such that:
\begin{equation*}
	\forall x_1, x_2 \in \mathbb{R}^m, ||f(x_1) - f(x_2)||_2 \leq L ||x_1 - x_2||_2
\end{equation*}
where $||\cdot||$ is $L_2$-norm of the Hilbert space $ \mathbb{R}^m$. 
The smallest $L$ for which the inequality holds is called the \textit{Lipschitz constant} of $f$ and is denoted as $L(f)$.
Intuitively, a function with a smaller Lipschitz constant changes less rapidly than a function with a larger Lipschitz constant.

For a typical multi-layer feed-forward neural network, the Lipschitz constant $L$ can be computed as 
\begin{equation} \label{eq:lipschitzConst}
	L(f) = \prod_{i} L(\phi_i)
\end{equation}
by using the composition property \cite{gouk2021regularisation}, where $\phi_i$ can be an affine, activation, or pooling operation. In this work, we use fully connected neural networks with 1-Lipschitz element-wise ReLU activation. Thus, the Lipschitz constant in \eqref{eq:lipschitzConst} can be further reduced to 
\begin{equation} \label{eq:lipschitzConst2}
	L(f) = \prod_{k=1}^K ||W_k||_2 = \prod_{k=1}^K \sigma_{\max}(W_k)
\end{equation}
where $W_k$ is the weight matrix of $k$th layer, and the singular value can be computed efficiently using the power method up to a given precision \cite{yoshida2017spectral}.

In the implementation, we apply spectral normalization to actor networks as follows:
\begin{equation*}
	W_k \leftarrow \frac{\sqrt[K]{L} \cdot W_k}{||W_k||_2}, 
\end{equation*}
such that the Lipschitz constant of the actor network is upper bounded by $L$, that is, $||f(x+\delta x) - f(x)||_2$ is always bounded by $L||\delta x||_2$. 
It is remarked that $L$ may not necessarily be the smallest Lipschitz constant of the actor network, even though $\sqrt[K]{L}$ is the best Lipschitz constant of each layer \cite{virmaux2018lipschitz}. Notably, this method is computationally efficient and integrates seamlessly with actor-critic-based RL algorithms. In this study, we propose and utilise \textit{spectral-normalized MADDPG} algorithm with $L=2.5$.

\section{Experimental Validation}\label{sec:exp}

To verify the effectiveness and robustness of the proposed methods in target state estimation  (Section \ref{sec:estimation}) and pursuit control (Section \ref{sec:rlFramework}), we developed a demonstration prototype. This section outlines the experimental setup and results.

\subsection{Experiment Setup} 

The experiment architecture consists of three main components: AGVs, a motion capture system, and a ROS-enabled computer (see Fig.~\ref{fig:experiment results}a). 
The custom-built AGV (see Fig.~\ref{fig:experiment results}b) measures around $30\times20\times15$ cm, and is equipped with a LoRa communication module for receiving control commands. 
The AGVs are assumed to be equipped with a monocular camera with a narrow $30\degree$ field of view, and bearings are reconstructed from motion capture data with deliberately added noise. We opted not to include a real camera, as the primary focus is on validating the target estimation and control algorithms, which operate independently of the visual detection component. This methodology is consistent with established practices \cite{he2019trajectory, li2022three} and ensures effective and relevant validation of the proposed algorithms.
It is highlighted that the $30\degree$ FoV, being significantly narrower than that of commonly used cameras, introduces increased difficulty, further testing the robustness of the proposed approach. 
Finally, the ROS-enabled computer runs the target state estimation and pursuit control algorithms at $10$ Hz. This choice considers the typically low update rate of onboard vision systems and the limited computational capacity of onboard computers. The AGV's inner velocity control is managed by a PID controller, operating at 100 Hz.

The desired range is set as $r_d=0.75$ m. 
The initial target estimate is $\hat p_T=[0, 0, 0]^T$ m and $\hat v_T=[0,0,0]^T$ m/s. Notably, the proposed u-PLIF demonstrates robustness to these initializations, ensuring that variations in these values do not significantly affect the convergence of estimation errors. 
The process noise covariance is set as $Q=0.25 \It$, and measurement noise covariance is $\Sigma = 10^{-4}\It$.

\subsection{Experiment Result} \label{sec:expResult}
At the beginning of the experiment, their heading angles are intentionally adjusted to deliberately fail in detecting the target (see Fig.~\ref{fig:experiment results}c), allowing an assessment of the target search capability.
The complete trajectories of the three pursuer vehicles and the target vehicle, and its corresponding estimate are shown in Fig.~\ref{fig:experiment results}d. 
It is evident that despite the target being manually controlled to execute an irregular circular motion, the estimated trajectory closely follows its movement.
Additionally, since agent 1 operates in unicycle mode, its trajectory exhibits a zigzag pattern to effectively track the target.
For a more detailed examination of the time-dependent evolution of the pursuers' headings, the complete trajectory is partitioned into five segments, each spanning a duration of 6 seconds. This subdivision is exemplified in Fig.~\ref{fig:experiment results}d. These sub-figures vividly illustrate the effectiveness of both position and heading control employed by the pursuers in maintaining the desired range and lines of sight to the target.

Figure~\ref{fig:experiment results}e presents the time-dependent evolution of various important properties pertaining to the cooperative pursuer team, target estimate errors, and control commands.
It is evident that initially none of the pursuers detect the target. However, the number of pursuers detecting the target gradually increases to three by approximately 4.5 seconds. 
It is worth highlighting a brief interruption in this trend at around $t=24$ s, where the detection count temporarily drops to two. This transient decrease is attributed to a sudden maneuver executed by the target under human control. 
However, this decrease is quickly rectified, and the estimation error of the target is unaffected as there are still two pursuers detecting it.
Throughout the experiment, the range control error decreases from around $1.5$ m to a level below 0.3 m. 
Remarkably, the observability consistently attains and maintains the theoretical maximum value of 2.25, ensuring the accuracy of target state estimation.

\begin{figure}[t]
	\centering
	\includegraphics[width=0.75\linewidth]{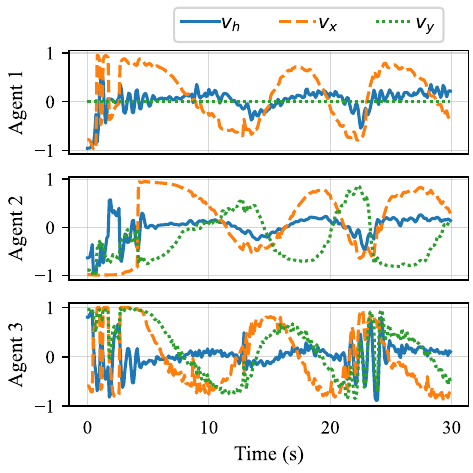}
	\caption{Compare time evolution of control commands in the experiment without target state estimation.}
	\label{fig:experiment results no noise}
\end{figure}

Figure~\ref{fig:experiment results}f provides insights into the behavior of position and velocity estimate errors. Notably, these errors decrease to nearly zero within approximately $1$ s. 
This behavior can be attributed to the capabilities of the cooperative u-PLIF approach. Once at least two bearing measurements are available, the target state estimate becomes highly reliable. Following convergence, the position estimate error remains within $0.02$ m, while the velocity estimate error stays within $0.2$ m/s.
The control commands generated by the policy networks are displayed in Fig.~\ref{fig:experiment results}g. The observed signal jitter primarily arises from noisy bearing measurements, as the control commands depend on these observations. 
To verify this, an additional experiment is conducted wherein the entire system operates without the target state estimation module. 
The results of this experiment are depicted in Fig.~\ref{fig:experiment results no noise}, demonstrating noticeably smoother control commands. This further confirms the effectiveness of the Lipschitz-constrained policy networks proposed in Section \ref{sec:spectralNormalization}.

\section{Discussion and Conclusion}
Unlike most MARL research, which primarily emphasizes learning efficiency, the significance of this work lies in its focus on delivering a fully integrated solution for target pursuit in real-world multi-robot systems. 
The proposed pseudo-linear information filter demonstrates notable stability and robustness, with potential applications across  various robotic and guidance systems. Additionally, the developed MARL framework provides a promising approach in addressing the multi-robot pursuit-evasion problem in complex environments. The sim-to-real techniques introduced in this study may offer valuable insights for the deployment of RL policies.
Future research will focus on addressing more aggressive target tracking and incorporating model-based approaches to enhance both explainability and performance.

%==========Bibliography==========
\bibliographystyle{IEEEtran}
\bibliography{Bibliography}
%\bibliography{nameFull,Bibliography}



\addtolength{\textheight}{-12cm}   % This command serves to balance the column lengths

\end{document}